\documentclass[10pt,twocolumn]{article}
\usepackage{latex8}
\usepackage{times}

\usepackage{rotating}
\usepackage{balance}

\usepackage{latexsym}
\usepackage{amsmath}
\usepackage{amssymb}
\usepackage{algorithm}
\usepackage{subfigure}
\usepackage{amsbsy,amsmath}

\def\P{\hbox{$\mathcal{P}$}}

\def\H{\hbox{$\mathcal{H}$}}
\def\M{\hbox{$\mathcal{M}$}}
\newcommand{\bs}[1]{\boldsymbol{#1}}
\newcommand{\PP}{\bs{\mathcal{P}}}

\newcommand{\keyw}[1]{\hbox{\bf #1}}
\newcommand{\floor}[1]{{\lfloor{#1}\rfloor}}

\DeclareMathAlphabet{\mathscr}{OT1}{pzc}%
                                 {m}{it}

\usepackage{graphicx}
\usepackage[dvips]{epsfig}

\usepackage{listings}

\begin{document}
%
\pagestyle{empty}
%


\title{Robust and Tuneable Family\\of Gossiping Algorithms}

\author{Vincenzo De Florio and Chris Blondia\\
University of Antwerp\\
Department of Mathematics and Computer Science\\
Performance Analysis of Telecommunication Systems group\\
Middelheimlaan 1, 2020 Antwerp, Belgium \vspace*{3pt} \\
Interdisciplinary Institute for Broadband Technology (IBBT)\\
Gaston Crommenlaan 8, 9050 Ghent-Ledeberg, Belgium}

\maketitle

\begin{abstract}
	We present a family of gossiping algorithms whose members share the same
structure though they vary their performance in function of a combinatorial parameter.
We show that such parameter may be considered as a ``knob'' controlling
the amount of communication parallelism characterizing the algorithms.
After this we introduce procedures to operate the knob and choose parameters
matching the amount of communication channels currently
provided by the available communication system(s). In so doing we
provide a robust mechanism to tune the production of requests for communication
after the current operational conditions of the consumers of such
requests. This can be used to achieve high performance
and programmatic avoidance of undesirable events such as message collisions.
\end{abstract}

\section{Introduction}
The main character in this text is a family
of algorithms for distributed gossiping whose members differ in
the strategy adopted to discipline the right to transmit. That strategy
can be expressed as a permutation of the indices of the participants.
In~\cite{DeFl06b} a formal model for this family
of algorithms was introduced and the performance of some of its members
was analyzed.
In the cited paper in particular it was shown how the choice of the structure of the permutations controlling
these algorithms
translates in different requirements on the underlying communication system---namely,
different amounts of concurrent send and receive requests.

The focus of this paper is not on the functional properties of the gossiping algorithms but rather
on the non-functional characteristics exhibited by them with the change of the adopted permutations.
Building on top of our past research, here we first show that by assigning different classes of
permutations to the participants
it is possible to scale dynamically the amount of communication requests triggered by the execution
of our algorithm. In other words, this enables the expression of a spectrum of codes each characterized by
a different algorithmic parallelism.

Secondly, we show here how this can be used to reach
an optimal match with the contextual (that is, physical) parallelism
provided by the deployment platform and networks.
Such an optimal tuning allows the avoidance of 
shortcoming and excess of algorithmic parallelism. While in the former case
one would under-utilize the available resources, in the latter case one would
issue a number of requests higher than the available communication resources,
which could lead to undesirable conditions such as 
overloading of request queues and packet collisions.

In what follows we first define our family of algorithms and concisely recall
its characteristics in Sect.~\ref{s:algo}.
Next, in Sect.~\ref{s:knob}, we introduce hybrid gossiping---a strategy to tune the
algorithmic parallelism 
in function of the physical parallelism characterizing the current context.
Section~\ref{s:mape} explains how that strategy can be used to design
autonomic evolution engines exploiting hybrid gossiping.
State of the art is then briefly summarized in Sect.~\ref{s:sota}.
Our conclusions follow in Sect.~\ref{s:conc}.

\section{A Family of Gossiping Algorithms}\label{s:algo}

In this section we recall the main features of a family of gossiping algorithms
firstly introduced in~\cite{DeFl06b}.
We refer the reader to the cited paper for a thorough discussion of the features of those algorithms. 
Introductions to gossiping,
which may be concisely defined as all-to-all pairwise inter-process communication,
may be found e.g. in~\cite{Goss,BGRV98,Gon03}.

In what follows we define a formal model for such family of
algorithms and we highlight the characteristics of two of its members.

\subsection{Formal Model}
Let $t$ represent time and $N>0$ be an integer.
We shall consider a set of $N+1$ communicating processes.
We assume that such processes may be uniquely identified via 
integers in $\{0,\dots,N\}$. Processes are deployed in
processing nodes linked together via one or more communication networks.
We shall refer to the set of nodes and networks as to ``the system''.
Nodes are equipped with a limited
number of communication ports. Likewise, networks provide a limited number
of independent full-duplex point-to-point communication lines.
At any time $t$ a new communication can only be initiated if a free port and a free
line are available in the system. If that is not the case, the requesting process is put in a wait state.
Due to resource competition, the number of ports and that of lines vary dynamically.
Depending on the available ports and lines, at any time $t$ 
at most $\mathcal{N}(t)$ send and at most $\mathcal{N}(t)$ receive requests
may be allowed to execute.
Communication is synchronous and blocking.


Processes own some local data they need to share
(for instance, to execute a voting algorithm as in~\cite{DeDL98e} or~\cite{BDB11a}).
In order to do so, each process broadcasts its local data
to all the others through multiple consecutive send requests,
and receives the $N$ data items owned by its fellows via
multiple consecutive receive requests.
A discrete time model is assumed---events occur at discrete time
steps, and during any time step any process can be involved in only
one such event.
More specifically, on a given time step $t$ process $i$ may be:
  \begin{enumerate}
    \item sending a message to process $j, j\neq i$; 
    this is represented as $i\, S^t j$;
    \item receiving a message from process $j, j\neq i$; 
    this is shown as $i\, R^t j$;
    \item blocked, waiting for messages to be received from any process;
    symbol ``$\curvearrowleft$'' will be used to mean this case;
    \item blocked, waiting for a message to be sent, i.e. for an addressee
    to enter the receiving state.
    Symbol ``$\curvearrowright$'' will be used for this.
\end{enumerate}

A \emph{slot\/} is defined as a process' temporal ``window'' one time step long.
On each given time step $t$, $N+1$ slots are available within the system.
Process $i$ \emph{makes use\/} of slot $t$
if and only if \( \exists \, j \, (i\, S^t j \vee i\, R^t j) \);
on the contrary, process $i$ is said to \emph{waste\/} slot $t$.

By the term ``run'' we shall refer in what follows to the
the collection of slots required to execute the above algorithm on a
given system, as well as to the values of the events corresponding to
those slots.

Let us define the following four state templates:

\begin{description}
\item[$W\!R$ state.]
A process is in state $W\!R_j$ if it is waiting for the arrival of a message
from process $j$. Where the subscript is not important it will be omitted.
Once in $W\!R$, a process stays there for one or more time steps, corresponding to the same number
of ``$\curvearrowleft$'' actions.

\item[$S$ state.]
A process $i$ is in state $S_j$ when it is sending a message to process $j$.
After one time step $i$ leaves state $S_j$.
This corresponds to one $i\, S^t j$ action.

\item[$W\!S$ state.]
A process $i$ waiting to send process $j$ its message is said to be in
state $W\!S_j$. Where the subscript is not important it will be omitted.
Once in $W\!S_j$, process $i$ stays there for one or more time steps, corresponding to the same number
of ``$\curvearrowright$'' actions.

\item[$R$ state.]
When process $i$ is receiving a message from process $j$, it is said to be in
state $R_j$. This state transition also lasts one time step and corresponds to action
$i\, R^t j$.
\end{description}

Let
${\P}_1,\dots,{\P}_N$ represent a permutation of
integers $0,\dots,i-1,i+1,\dots,N$.
Then the above state templates can be used to compose $N+1$
finite state automata as described in Algorithm~1.

\begin{algorithm*}
\begin{center}\parbox{0cm}{\begin{tabbing}
xx \= xx \= xx \= xx \= xx \= xx \= xx \= xx \= xx \= xx \= xx \= xx \= xxxxx \= xx \kill
	 \> \> \textbf{Algorithm~1: Compose the FSA solving gossiping}\\
	 \> \> \> \> \> \> \> \textbf{for process $i\in\{0,\dots,N\}$}\\
         \> \> \emph{Input:\/} $A \equiv (i, N, \P)$\\
         \> \> \emph{Output:\/} $\mathrm{FSA}(A)$\\
{\bf  1} \> \> \keyw{begin}\\
         \> \> \> /* emit the initial state */ \\
{\bf  2} \> \> \> $\mathrm{FSA}(A) := \mathrm{START}$\\
{\bf  3} \> \> \> \keyw{for} $j := 0$ \keyw{to} $i-1$ \keyw{do} \\
         \> \> \> \> /* operator ``$\leftarrow$'' appends a new state to the FSA */ \\
{\bf  4} \> \> \> \> $\mathrm{FSA}(A) \leftarrow W\!R$ \\
{\bf  5} \> \> \> \> $\mathrm{FSA}(A) \leftarrow R$ \\
{\bf  6} \> \> \> \keyw{enddo}\\
{\bf  7} \> \> \> \keyw{for} $j := 1$ \keyw{to} $N$ \keyw{do} \\
{\bf  8} \> \> \> \> $\mathrm{FSA}(A) \leftarrow W\!S_{{\P}_j}$\\
{\bf  9} \> \> \> \> $\mathrm{FSA}(A) \leftarrow S_{{\P}_j}$\\
{\bf 10} \> \> \> \keyw{enddo}\\
{\bf 11} \> \> \> \keyw{for} $j := i+1$ \keyw{to} $N$ \keyw{do}\\
{\bf 12} \> \> \> \> $\mathrm{FSA}(A) \leftarrow W\!R$\\
{\bf 13} \> \> \> \> $\mathrm{FSA}(A) \leftarrow R$\\
{\bf 14} \> \> \> \keyw{enddo}\\
         \> \> \> /* emit the final state */ \\
{\bf 15} \> \> \> $\mathrm{FSA}(A) \leftarrow \mathrm{STOP}$\\
{\bf 16} \> \> \keyw{end}.
\end{tabbing}}
\end{center}
\end{algorithm*}

\begin{figure*}[t]
\centerline{\psfig{figure=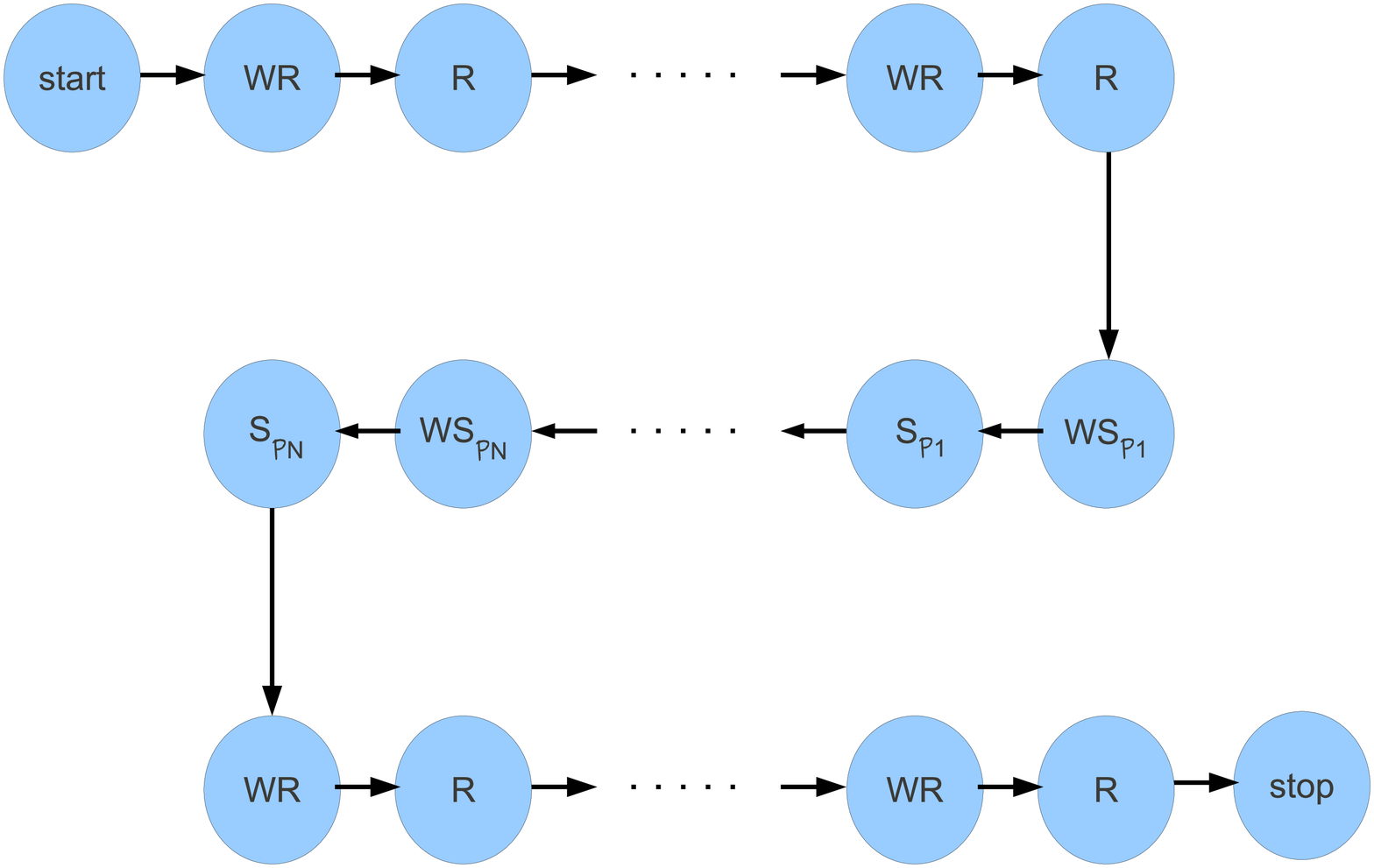,width=0.86\textwidth}}
\caption{The state diagram of the FSA run by process $i$. The first
row consists of $i$ \( (W\!R, R) \) couples.
$({\P}_1,\dots,{\P}_N)$ represents a permutation of the $N$ integers
$0,\dots,i-1,i+1,\dots,N$.
The last row contains $N-i$ \( (W\!R, R) \) couples.}
\label{fsa}
\end{figure*}

Figure~\ref{fsa} shows the finite state automaton that solves distributed
gossiping for process $i$, which we obtained by executing Alg.~1.
The first row is the condition that has to be reached before
process $i$ is allowed to begin its broadcast: a series
of $i$ \( (W\!R, R) \) couples.

Once process $i$ has successfully received $i$ messages, it acquires
the right to broadcast. Broadcasting is performed according to the rule expressed in
the second row of Fig.~\ref{fsa}: process $i$ orderly sends its message to
its fellows, the $j$-th message being sent to process ${\P}_j$.

The third row of Fig.~\ref{fsa} instructs the reception of the remaining $N-i$ messages,
which is coded as a sequence of $N-i$ \( (W\!R, R) \) couples.

In~\cite{DeFl06b} it was shown how, irrespective of the value of $\P$, such FSA's
implement a distributed deadlock-free gossiping algorithm.
As intuition may suggest, the choice of which permutation to use has
indeed a deep impact
on the overall performance of the algorithm---together
with the physical characteristics of the system. In fact,
different permutations translate in different amounts of
communication parallelism; when such algorithmic parallelism is
backed up by contextual parallelism---that is, by
a sufficiently large number of
independent communication ports and lines in the system,
modeled as dynamic system $\mathcal{N}(t)$---then there is
an optimal match between the algorithm and the deployment
platform.

In order to evaluate the above impact
we shall make use of the following ``quality metrics'':
\begin{description}
\item[Average slot utilization.]
This is the average number of used slots per time step in a given run.
It will be indicated as $\mu_N$, or simply as $\mu$.
$\mu$ can be interpreted as
the average degree of parallelism expressed by the algorithm---hence it will
be referred to also as the ``algorithmic parallelism''.
$\mu$ can take any real value in $[0, N+1]$.


\item[Length.]
This is the number of time steps in a run. It represents a measure
of the time needed for the distributed algorithm to complete.
$\lambda_N$, or more simply $\lambda$, will be used for lengths.
\end{description}


For any time step $t$, we shall call $\nu_t$ as the number of slots that
were used during $t$.
The $\lambda$-tuple
\( \vec{\nu} = [ \nu_1, \nu_2, \dots, \nu_\lambda ],\)
orderly encoding the number of used slots for
each time step in a run, shall be called ``utilization string.''

In~\cite{DeFl06b} several cases of $\P$ were introduced and discussed.
In particular in the above cited reference it was shown how varying the structure of $\P$
produces quite different values of $\mu$ %
and $\lambda$.
This fact, coupled with physical constraints of the system
as modeled by $\mathcal{N}(t)$,
determine the overall performance of our algorithm.

In what follows we focus on two particular cases representing 
the minimal and maximal algorithmic parallelism.

\subsection{Identity Permutation}

As a first case, 
let $\P$ be the identity permutation:
\begin{equation}\label{zeroperm}
\binom{0,\dots,i-1,i+1,\dots,N}{0,\dots,i-1,i+1,\dots,N},
\end{equation}
i.e., in cycle notation~\cite{Knu73a},
\(
\P = (0)\dots(i-1)(i+1)\dots(N).
\)

This means that, once process $i$ acquires the right to broadcast,
it first sends its message to process 0
(possibly having to wait for it to become available),
then it will do the same with process 1, and so
forth up to process $N$, obviously skipping itself. This is
shown in Table~\ref{run4} for $N=5$. In what follows we shall refer to tables
such as Table~\ref{run4} as to ``run-tables.''

\begin{table*}[t]
\begin{center}
\begin{tabular}{l@{}|@{}c@{}c@{}c@{}c@{}c@{}c@{}c@{}c@{}c@{}c@{}c@{}c@{}c@{}c@{}c@{}c@{}c@{}c@{}c@{}c@{}c@{}c@{}c@{}c@{}c@{}c@{}}
$\stackrel{\hbox{\sffamily\scriptsize id}}{\downarrow}\stackrel{\begin{sideways}\hbox{\sffamily\scriptsize step}\end{sideways}}{\rightarrow}$&\tiny1&\tiny2&\tiny3&\tiny4&\tiny5&\tiny6&\tiny7&\tiny8&\tiny9&\tiny10&\tiny11&\tiny12&\tiny13&\tiny14&\tiny15&\tiny16&\tiny17&\tiny18&\tiny19&\tiny20&\tiny21&\tiny22&\tiny23&\tiny24&\tiny25&\tiny26\\\hline
\sf 0&$S_{1}$&$S_{2}$&$S_{3}$&$S_{4}$&$S_{5}$&$R_{1}$&$\curvearrowleft$&$R_{2}$&$\curvearrowleft$&$\curvearrowleft$&$\curvearrowleft$&$\curvearrowleft$&$R_{3}$&$\curvearrowleft$&$\curvearrowleft$&$\curvearrowleft$&$R_{4}$&$\curvearrowleft$&$\curvearrowleft$&$\curvearrowleft$&$\curvearrowleft$&$R_{5}$&$\curvearrowleft$&$\curvearrowleft$&$\curvearrowleft$&$\curvearrowleft$\\
\sf 1&$R_{0}$&$\curvearrowright$&$\curvearrowright$&$\curvearrowright$&$\curvearrowright$&$S_{0}$&$S_{2}$&$S_{3}$&$S_{4}$&$S_{5}$&$R_{2}$&$\curvearrowleft$&$\curvearrowleft$&$R_{3}$&$\curvearrowleft$&$\curvearrowleft$&$\curvearrowleft$&$R_{4}$&$\curvearrowleft$&$\curvearrowleft$&$\curvearrowleft$&$\curvearrowleft$&$R_{5}$&$\curvearrowleft$&$\curvearrowleft$&$\curvearrowleft$\\
\sf 2&$\curvearrowleft$&$R_{0}$&$\curvearrowleft$&$\curvearrowleft$&$\curvearrowleft$&$\curvearrowleft$&$R_{1}$&$S_{0}$&$\curvearrowright$&$\curvearrowright$&$S_{1}$&$S_{3}$&$S_{4}$&$S_{5}$&$R_{3}$&$\curvearrowleft$&$\curvearrowleft$&$\curvearrowleft$&$R_{4}$&$\curvearrowleft$&$\curvearrowleft$&$\curvearrowleft$&$\curvearrowleft$&$R_{5}$&$\curvearrowleft$&$\curvearrowleft$\\
\sf 3&$\curvearrowleft$&$\curvearrowleft$&$R_{0}$&$\curvearrowleft$&$\curvearrowleft$&$\curvearrowleft$&$\curvearrowleft$&$R_{1}$&$\curvearrowleft$&$\curvearrowleft$&$\curvearrowleft$&$R_{2}$&$S_{0}$&$S_{1}$&$S_{2}$&$S_{4}$&$S_{5}$&$\curvearrowleft$&$\curvearrowleft$&$R_{4}$&$\curvearrowleft$&$\curvearrowleft$&$\curvearrowleft$&$\curvearrowleft$&$R_{5}$&$\curvearrowleft$\\
\sf 4&$\curvearrowleft$&$\curvearrowleft$&$\curvearrowleft$&$R_{0}$&$\curvearrowleft$&$\curvearrowleft$&$\curvearrowleft$&$\curvearrowleft$&$R_{1}$&$\curvearrowleft$&$\curvearrowleft$&$\curvearrowleft$&$R_{2}$&$\curvearrowleft$&$\curvearrowleft$&$R_{3}$&$S_{0}$&$S_{1}$&$S_{2}$&$S_{3}$&$S_{5}$&$\curvearrowleft$&$\curvearrowleft$&$\curvearrowleft$&$\curvearrowleft$&$R_{5}$\\
\sf 5&$\curvearrowleft$&$\curvearrowleft$&$\curvearrowleft$&$\curvearrowleft$&$R_{0}$&$\curvearrowleft$&$\curvearrowleft$&$\curvearrowleft$&$\curvearrowleft$&$R_{1}$&$\curvearrowleft$&$\curvearrowleft$&$\curvearrowleft$&$R_{2}$&$\curvearrowleft$&$\curvearrowleft$&$R_{3}$&$\curvearrowleft$&$\curvearrowleft$&$\curvearrowleft$&$R_{4}$&$S_{0}$&$S_{1}$&$S_{2}$&$S_{3}$&$S_{4}$\\
\hline
$\stackrel{\hbox{\sffamily\scriptsize used}}{\rightarrow}$&2&2&2&2&2&2&2&4&2&2&2&2&4&4&2&2&4&2&2&2&2&2&2&2&2&2
\end{tabular}
\end{center}
\caption{Run ($N=5$) for $\P$ equal to the identity permutation.
The {\sf step} row represents time steps. {\sf Id}'s identify
processes. $\vec\nu$ is the utilization string.
In this case
$\mu$ (that is, algorithmic parallelism) is about 2.3077 slots,
and length $\lambda=26$.
Note that, if the slot is a used one, then entry $(i,t)={\mathcal R}_j$ of this matrix
is action $i\, {\mathcal R}^t j$.}\label{run4}
\end{table*}

In~\cite{DeFl06b} it was shown how
it is possible to characterize some properties of the
quality metrics of the member corresponding to the
identity permutation. Among such properties particularly useful here are
the following two ones:
\begin{itemize}
\item The algorithm makes use of $\hbox{O}(N^2)$ time---more precisely,
\(
	\lambda_N = \frac34N^2 + \frac54N + \frac12 \floor{N/2}.
\)
\item The asymptotic value of algorithmic parallelism, that is
\(
	\lim_{k\rightarrow\infty}\mu_k,
\)
equals $\frac83$.
\end{itemize}


\subsection{Pipelined Permutation}\label{tao-pb}
We now consider a second case---the one corresponding to
permutation

\begin{equation}\label{pipe}
\binom{0,\dots,i-1,i+1,\dots,N}%
      {i+1,\dots,N,0,\dots,i-1}.
\end{equation}

Note how permutation~(\ref{pipe}) is equivalent to $i$ 
cyclic logical left shifts of the identity permutation.

When $\P$ is as in (\ref{pipe}), then
process $i$ first sends its message to process $i+1$,
then to process $i+2$, and so
on until it reaches process $N$. After that, $i$ wraps around
and sends from process 0 to process $i-1$.
This is shown in Table~\ref{run9} for $N=8$.
As can be seen from that table, (\ref{pipe}) maximally overlaps the processes'
broadcast sessions
the same way as machine instructions are being overlapped in
pipelined microprocessors~\cite{He06}. This similarity brought
to the name of ``pipelined permutation'' for (\ref{pipe})~\cite{DeFl06b}. 


\begin{table*}[t]
\begin{center}
\begin{tabular}{l@{}|@{}c@{}c@{}c@{}c@{}c@{}c@{}c@{}c@{}c@{}c@{}c@{}c@{}c@{}c@{}c@{}c@{}c@{}c@{}c@{}c@{}c@{}c@{}c@{}c@{}}
$\stackrel{\hbox{\sffamily\scriptsize id}}{\downarrow}$ $\stackrel{\begin{sideways}\hbox{\sffamily\scriptsize step}\end{sideways}}{\rightarrow}$&\tiny1&\tiny2&\tiny3&\tiny4&\tiny5&\tiny6&\tiny7&\tiny8&\tiny9&\tiny10&\tiny11&\tiny12&\tiny13&\tiny14&\tiny15&\tiny16&\tiny17&\tiny18&\tiny19&\tiny20&\tiny21&\tiny22&\tiny23&\tiny24\\ \hline
\sf 0&$S_{1}$&$S_{2}$&$S_{3}$&$S_{4}$&$S_{5}$&$S_{6}$&$S_{7}$&$S_{8}$&$\curvearrowleft$&$R_{1}$&$R_{2}$&$R_{3}$&$R_{4}$&$R_{5}$&$R_{6}$&$R_{7}$&$R_{8}$&$\curvearrowleft$&$\curvearrowleft$&$\curvearrowleft$&$\curvearrowleft$&$\curvearrowleft$&$\curvearrowleft$&$\curvearrowleft$\\
\sf 1&$R_{0}$&$\curvearrowright$&$S_{2}$&$S_{3}$&$S_{4}$&$S_{5}$&$S_{6}$&$S_{7}$&$S_{8}$&$S_{0}$&$\curvearrowleft$&$R_{2}$&$R_{3}$&$R_{4}$&$R_{5}$&$R_{6}$&$R_{7}$&$R_{8}$&$\curvearrowleft$&$\curvearrowleft$&$\curvearrowleft$&$\curvearrowleft$&$\curvearrowleft$&$\curvearrowleft$\\
\sf 2&$\curvearrowleft$&$R_{0}$&$R_{1}$&$\curvearrowright$&$S_{3}$&$S_{4}$&$S_{5}$&$S_{6}$&$S_{7}$&$S_{8}$&$S_{0}$&$S_{1}$&$\curvearrowleft$&$R_{3}$&$R_{4}$&$R_{5}$&$R_{6}$&$R_{7}$&$R_{8}$&$\curvearrowleft$&$\curvearrowleft$&$\curvearrowleft$&$\curvearrowleft$&$\curvearrowleft$\\
\sf 3&$\curvearrowleft$&$\curvearrowleft$&$R_{0}$&$R_{1}$&$R_{2}$&$\curvearrowright$&$S_{4}$&$S_{5}$&$S_{6}$&$S_{7}$&$S_{8}$&$S_{0}$&$S_{1}$&$S_{2}$&$\curvearrowleft$&$R_{4}$&$R_{5}$&$R_{6}$&$R_{7}$&$R_{8}$&$\curvearrowleft$&$\curvearrowleft$&$\curvearrowleft$&$\curvearrowleft$\\
\sf 4&$\curvearrowleft$&$\curvearrowleft$&$\curvearrowleft$&$R_{0}$&$R_{1}$&$R_{2}$&$R_{3}$&$\curvearrowright$&$S_{5}$&$S_{6}$&$S_{7}$&$S_{8}$&$S_{0}$&$S_{1}$&$S_{2}$&$S_{3}$&$\curvearrowleft$&$R_{5}$&$R_{6}$&$R_{7}$&$R_{8}$&$\curvearrowleft$&$\curvearrowleft$&$\curvearrowleft$\\
\sf 5&$\curvearrowleft$&$\curvearrowleft$&$\curvearrowleft$&$\curvearrowleft$&$R_{0}$&$R_{1}$&$R_{2}$&$R_{3}$&$R_{4}$&$\curvearrowright$&$S_{6}$&$S_{7}$&$S_{8}$&$S_{0}$&$S_{1}$&$S_{2}$&$S_{3}$&$S_{4}$&$\curvearrowleft$&$R_{6}$&$R_{7}$&$R_{8}$&$\curvearrowleft$&$\curvearrowleft$\\
\sf 6&$\curvearrowleft$&$\curvearrowleft$&$\curvearrowleft$&$\curvearrowleft$&$\curvearrowleft$&$R_{0}$&$R_{1}$&$R_{2}$&$R_{3}$&$R_{4}$&$R_{5}$&$\curvearrowright$&$S_{7}$&$S_{8}$&$S_{0}$&$S_{1}$&$S_{2}$&$S_{3}$&$S_{4}$&$S_{5}$&$\curvearrowleft$&$R_{7}$&$R_{8}$&$\curvearrowleft$\\
\sf 7&$\curvearrowleft$&$\curvearrowleft$&$\curvearrowleft$&$\curvearrowleft$&$\curvearrowleft$&$\curvearrowleft$&$R_{0}$&$R_{1}$&$R_{2}$&$R_{3}$&$R_{4}$&$R_{5}$&$R_{6}$&$\curvearrowright$&$S_{8}$&$S_{0}$&$S_{1}$&$S_{2}$&$S_{3}$&$S_{4}$&$S_{5}$&$S_{6}$&$\curvearrowleft$&$R_{8}$\\
\sf 8&$\curvearrowleft$&$\curvearrowleft$&$\curvearrowleft$&$\curvearrowleft$&$\curvearrowleft$&$\curvearrowleft$&$\curvearrowleft$&$R_{0}$&$R_{1}$&$R_{2}$&$R_{3}$&$R_{4}$&$R_{5}$&$R_{6}$&$R_{7}$&$\curvearrowright$&$S_{0}$&$S_{1}$&$S_{2}$&$S_{3}$&$S_{4}$&$S_{5}$&$S_{6}$&$S_{7}$\\
\hline
$\stackrel{\hbox{\scriptsize $\vec{\nu}$}}{\rightarrow}$&2&2&4&4&6&6&8&8&8&8&8&8&8&8&8&8&8&8&6&6&4&4&2&2
\end{tabular}
\end{center}
\caption{Run-table for $N=8$ and the pipelined permutation.
In this case algorithmic parallelism is 6
and length is 24. Efficiency (channel exploitation) is of 6 slots out of 9, that is 66.67\%.}
\label{run9}
\end{table*}

In~\cite{DeFl06b} 
some of the quality metrics of the member corresponding to the
pipelined permutation were also characterized. In particular it was shown that:
\begin{itemize}
\item The algorithm makes use of $\hbox{O}(N)$ time, and more precisely
\(
	\lambda_N = 3N.
\)
\item Algorithmic parallelism linearly depends on the amount of involved processes:
\( \forall k>0: \mu_k = \frac23 (k+1).\)
\end{itemize}

\section{Tuning Algorithmic Parallelism through Hybrid Gossiping}\label{s:knob}
\begin{figure*}[t]
\centerline{\psfig{figure=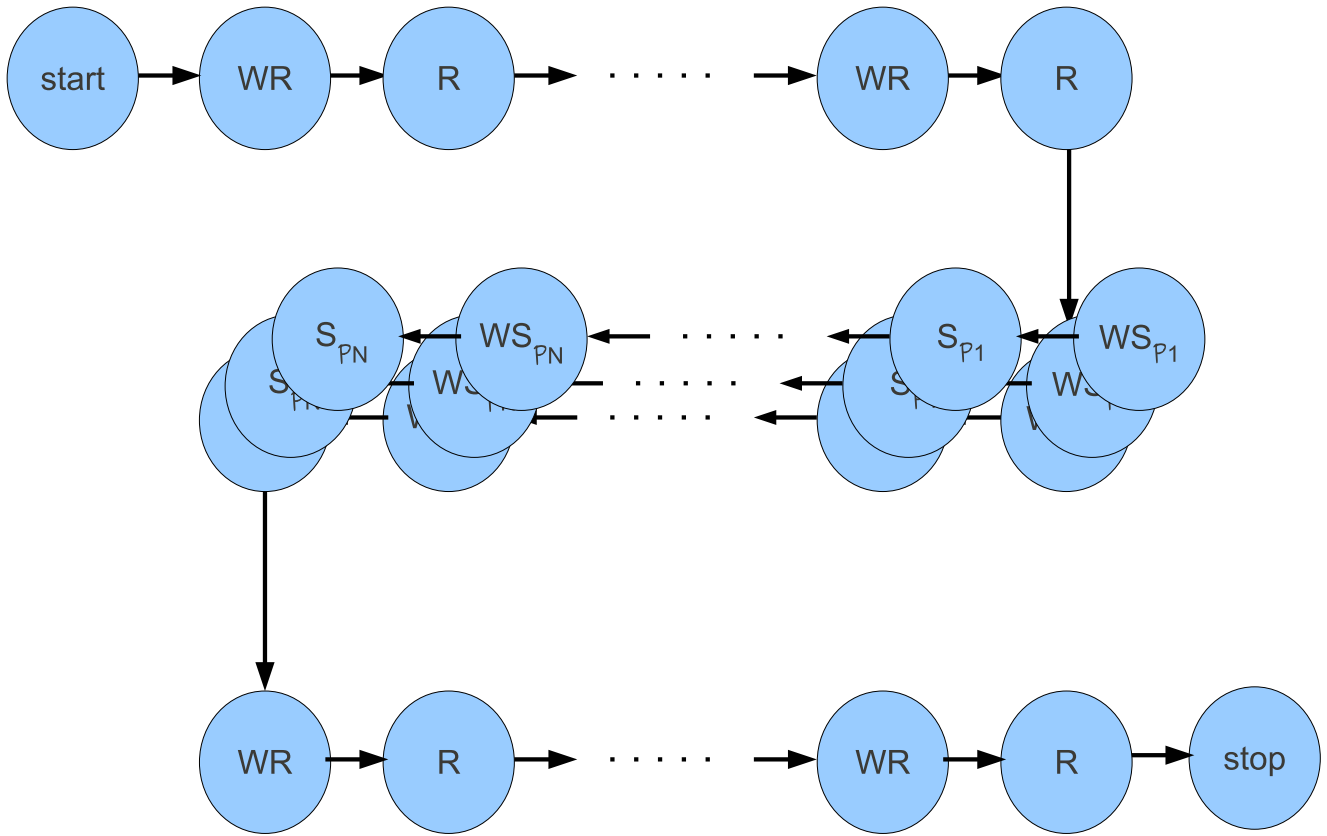,width=1.0\textwidth}}
\caption{Hybrid gossiping: processes operate using a permutation selected from two or more classes.}
\label{fsa2}
\end{figure*}

The two cases introduced in the previous section represent
two ``extremes'' in the spectrum of possible permutation structures:
the identity permutation and the pipelined permutation
respectively translate in very low and very high
algorithmic parallelism.
These emerging behaviors characterize \emph{homogeneous\/} gossiping---gossiping
that is in which all processes make use of the same permutation.
A useful property of our family of algorithms is that it also supports
\emph{hybrid\/} gossiping: in this case the gossiping processes make use
of a permutation selected (with some predefined logic) from two or more classes,
as depicted in Fig.~\ref{fsa2}. 
A noteworthy assignment logic is the one that schedules the pipelined permutation
to a certain percentage of the processes and the identity permutation to the
rest. By doing so we experimentally found that the ensuing algorithmic parallelism
grows after the percentage of pipelined permutations assigned to the processes.
In what follows we shall use symbol ``$\H_N$'' 
(or where possible without ambiguity, ``$\H$'') to refer to such percentage.
Figure~\ref{f:avg} shows values of $\mu$ (algorithmic parallelism)
for up to 50 participating processes when various percentages
are used. 

\begin{figure*}[t]
\centerline{\input average
}
\caption{Hybrid gossiping produces different amounts
of algorithmic parallelism depending on the scheduling distribution
of the pipelined and identity permutations.}
\label{f:avg}
\end{figure*}

\section{Autonomic System Evolution}\label{s:mape}
The ability to tune algorithmic parallelism paves the way to the definition
of autonomic procedures to evolve the system. In general such evolutions take the
form of a so-called ``MAPE'' adaptation loop~\cite{KeCh:2003}, where MAPE
stands for ``Monitor-Analyze-Plan-Execute''. In the case at hand,
\begin{description}
  \item[Monitor]
      signifies being able to estimate $\mathcal{N}(t)$, 
  \item[Analyze] means
      checking whether $\mu_N$ matches the estimated value of $\mathcal{N}(t)$,
  \item[Plan] is a ``meta-algorithm'' (also known as ``evolution engine''~\cite{DeF11a})
	  responsible for choosing how to evolve the system,
\item[Execute] is the execution of the meta-algorithm and the corresponding
      evolution of the managed system.
\end{description}

\begin{figure*}[t]
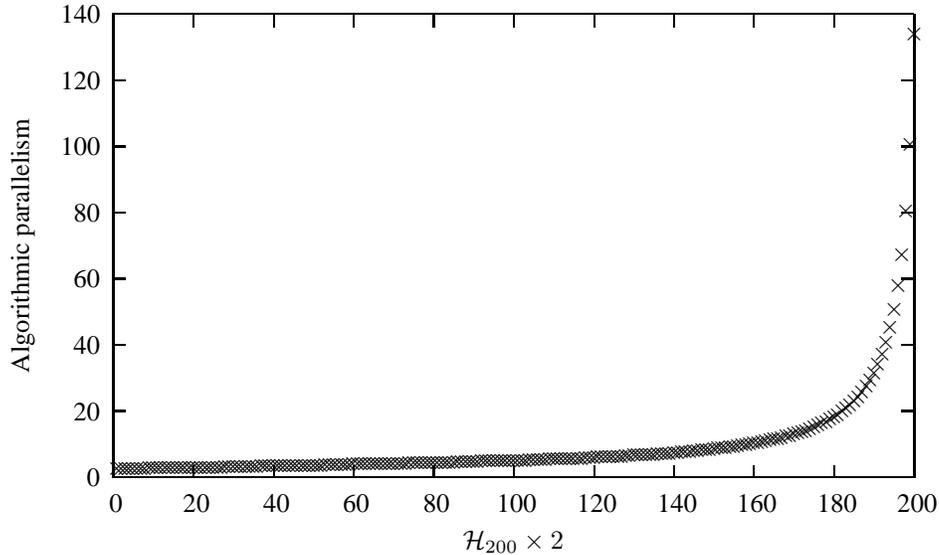

\centerline{\input h200
}
\caption{Algorithmic parallelism when $\H_{200}$ varies from 0.5\% to 100\% with step 0.5\%.}
\label{f:h200}
\end{figure*}

In what follows we assume the availability of a monitoring function called
\keyw{sense}. A reflective system such as the one introduced
in~\cite{De10} or~\cite{DB07d} could be used to provide transparent access
to the number of currently available ports and lines---that is, $\mathcal{N}(t)$.
The ``Analyze'' step is merely the assessment of
how close the current value of
$\H_N$ is to the estimated value of $\mathcal{N}(t)$. The system would evolve
only in case of two conditions: \emph{overshooting}, that is
a value of $\H_N$ overabundant with respect to that of $\mathcal{N}(t)$, and
\emph{undershooting}---namely, a value of $\H_N$ that would translate
in a sub-optimal exploitation of the available contextual parallelism.
In case the system would indeed require adaptation, several
meta-algorithms may be selected for the Planning step depending on e.g. the characteristic
of the mission and the system assumptions: in fact complex planning is likely
to call for non-negligible amounts of system resources, which could
interfere e.g. with real-time requirements. A possible cost-effective
solution for the meta-algorithm could then be to make use of
look-up tables providing for several values of $N$
the algorithmic parallelism corresponding to some sampling of $\H_N$.
Figure~\ref{f:h200} shows 200 samples of $\H_{200}$, which could be
computed off-line and stored in one such look-up table.
Algorithm~2 could then be used for the ``Execute'' step.
Alternatively, should performance penalties be deemed preferable
to the memory penalty to store the look-up table,
one could compute the curve best fitting the sampling of $\H_N$ at run-time.

\begin{algorithm*}
\begin{center}\parbox{0cm}{\begin{tabbing}%
xx \= xx \= xx \= xx \= xx \= xx \= xx \= xx \= xx \= xx \= xx \= xx \= xxxxx \= xx \kill
\> \> \textbf{Algorithm~2: Tune algorithmic parallelism}\\
\> \> \> \> \> \> \> \textbf{after contextual parallelism}\\
\> \> \emph{Input:\/} $\mathcal{N}(t), N$, look-up table $\M_N$; \emph{Output:\/} $\H_{\hbox{best}}$\\
{\bf  1} \> \> \keyw{begin}\\
         \> \> \> /* $cp$ holds the current contextual parallelism */\\
         \> \> \> /* Function \keyw{now} returns current time */\\
{\bf  2} \> \> \> $cp \leftarrow \keyw{sense}(\mathcal{N}(\keyw{now}()))$\\
{\bf  3} \> \> \> $\H_{\hbox{best}} \leftarrow \min \{ h : \M_N(h) \ge cp \}$\\
{\bf  4} \> \> \> \keyw{return} $\H_{\hbox{best}}$ \\
{\bf  5} \> \> \keyw{end}.
\end{tabbing}}
\end{center}
\end{algorithm*}

Another possibility is for instance the one described in
Algorithm~3.
In this case we define $\M_N$ as an associative map, that is,
a growing set of domain-to-value associations
that link known values of $\H_N$ to corresponding known values of $\mu_N$.
A system such as the one described in~\cite{Dev27} could implement such a map.
The specific difference with respect to a look-up table lies in the
dynamic nature of $\M_N$,
as it is possible to add new associations
to it at all times. It is assumed that $\M_N$ is initialized at least
with the associations corresponding
to the identity and the pipelined permutations.

\begin{figure*}[t]
\centerline{\psfig{figure=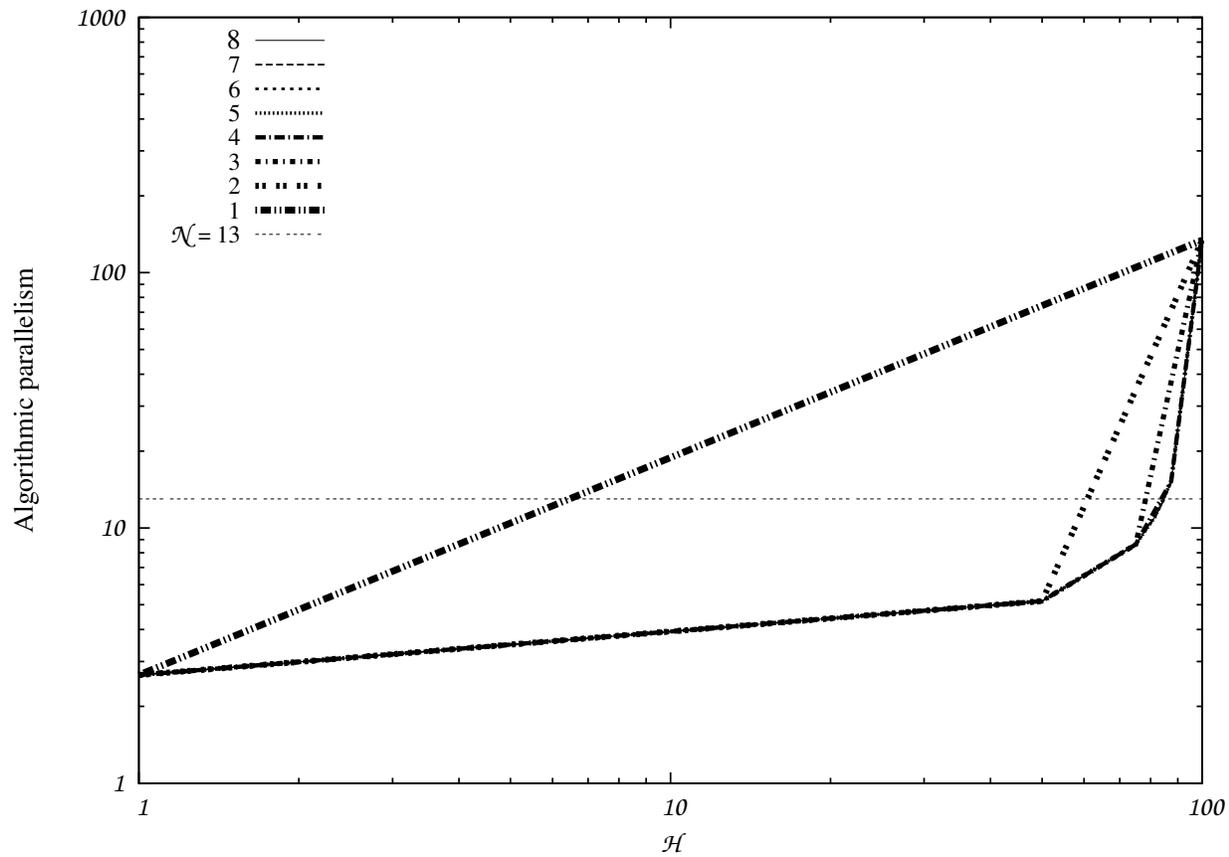,width=0.950\textwidth}}
\caption{A graphical representation of Algorithm~3 when $N=200$ and
$\forall t: \mathcal{N}(t)=13$. Logarithmic scales are used.}
\label{f:2batw}
\end{figure*}

The strategy followed in this case is to return a best match with
the entries currently available in $\M_N$, and
to add new entries to refine dichotomically the list.
The current linear strategy may be replaced by a more efficient one
designed after the non-linear nature of $\mu$.
Figure~\ref{f:2batw} graphically depicts the working of
Algorithm~3 under the following conditions: $N=200$; $\M_{200}$ initially includes only
the identity and pipelined permutations; $\mathcal{N}(t)=13$ for all values of $t$.
At the eighth iteration $|\M_{200}|=9$ and the selected
best value for $\mu$ is 13.11.

\begin{algorithm*}[t]
\begin{center}\parbox{0cm}{\begin{tabbing}
xx \= xx \= xx \= xx \= xx \= xx \= xx \= xx \= xx \= xx \= xx \= xx \= xxxxx \= xx \kill
\> \> \textbf{Algorithm~3: Tune algorithmic parallelism}\\
\> \> \> \> \> \> \> \textbf{after contextual parallelism}\\
         \> \> \emph{Input:\/} $\mathcal{N}(t), N$, associative map $\M_N$;
	 \emph{Output:\/} $\H_{\hbox{best}}$\\
{\bf  1} \> \> \keyw{begin}\\
{\bf  2} \> \> \> $cp \leftarrow \keyw{sense}(\mathcal{N}(\keyw{now}()))$\\
{\bf  3} \> \> \> $\H_{\hbox{best}} \leftarrow \min \{ h : \M_N(h) \ge cp \}$\\
{\bf  4} \> \> \> \keyw{if} $\M_N( \H_{\hbox{best}} ) = cp$ \keyw{then} \keyw{return} $\H_{\hbox{best}}$ \keyw{fi} \\
         \> \> \> /* $\M_N( \H_{\hbox{best}} ) > cp$ */ \\
{\bf  5} \> \> \> $sb \leftarrow 
			\max \{ h : \M_N(h) < \M_N(\H_{\hbox{best}}) \}$\\
{\bf  6} \> \> \> $\H_{\hbox{best}} \leftarrow (\M_N(\H_{\hbox{best}}) - \M_N(sb))/2$\\
\> \> \> /* $\keyw{compute}_\mu$ simulates a run */\\
\> \> \> /* and computes the corresponding $\mu$ */\\
{\bf  7} \> \> \> $\mu \leftarrow \keyw{compute}_\mu(\H_{\hbox{best}})$\\
{\bf  8} \> \> \> $\M_N \leftarrow \M_N \cup \{ \H_{\hbox{best}} \to \mu \}$\\
{\bf  9} \> \> \> \keyw{return} $\H_{\hbox{best}}$\\
{\bf 10} \> \> \keyw{end}.
\end{tabbing}}
\end{center}
\end{algorithm*}

\section{State of the Art}\label{s:sota}
In this section we briefly report on methods and strategies to tune
the characteristics of a software system so as to achieve
high performance e.g. through the expression and exploitation of
algorithmic parallelism. 

Concerns such as this
normally are not localized in a single physical module (e.g. a function) or
logical component (e.g., a formal parameter, as it is the case for our gossiping
algorithms);
on the contrary, they often span through several modules and require the
joint evolution of multiple correlated variables. In other words,
the expression of parallelism is a typical \emph{cross-cutting concern}.
A typical way to tackle such concerns is through the usage of
Aspect Oriented Programming (AOP)~\cite{KLM97}.
In AOP the source core consists of two separate ``blocks'':
the functional code dealing with the business logic and the aspect code
to express one or more cross-cutting concerns. The actual source code
is the result of a merging process called weaving where the functional code
is rearranged, instrumented, and patched, according to what
prescribed in the aspect code. This allows software systems to be
effectively evolved so as to maximize one or more target
concerns---including e.g. the expression of algorithmic parallelism~\cite{Sub:2005}.
Widely used in industry and academia, AOP
proved in many cases to be able to manage effectively the complexity of software
evolution.

The general problem of guaranteeing the emergence of certain expected features
or behaviors in a system---software or otherwise---was termed by Jen~\cite{Jen04:Jen}
as ``robust evolvability''. In the cited paper the author defines
evolvability as an entity's ability to
``alter their structure or function so as to adapt to
changing circumstances'' and discusses
a system's capability to retain certain characteristics of interests
(e.g. maximizing algorithmic parallelism)
despite changes in its composition and deployment environment.
Such capability is called by Jen as robustness:
feature persistence under specified and unforeseen perturbations, obtained by
switching among multiple stra\-te\-gic options such that
those changes are dynamically tolerated or even exploited.
Interestingly enough, Jen distinguishes two classes of evolvable systems.
\begin{itemize}
\item Phenotypically plastic systems. Such systems retain their
structure and organization throughout adaptations
and only achieve evolution by switching
among a few, preordained, structurally equivalent configurations that depend on
some internal parameter. Obviously this is the case for our gossiping algorithms.
\item Phenotypically dynamic systems programming system,
which are able to assume
different structures and organizations by mutating the topology,
the role, and the number of their components. An example of this is given
by AOP systems.
\end{itemize}
A deeper discussion and some examples of systems matching the above classes may be found
e.g. in~\cite{DeF11a}.

We believe as worth mentioning here the case of FFTW, an evolvable software system that tunes its
logics so as to maximize performance on a given target platform. FFTW (whose name stands for
``Fastest Fourier Transform in the West'')
is a code generator for Fast Fourier Transforms that defines and assembles
blocks of C code that optimally solve FFT sub-problems
on a given machine~\cite{Fri04}.

Finally we observe how nowadays it is common practice designing software for \emph{families\/} of
target platforms, which selectively enable or disable target-specific
optimizations. One such software is the mplayer video
player~\cite{Mplayer}, which explicitly states such
optimizations with messages such as
``Using optimized IMDCT transform'' or ``Using MMX optimized resampler.''
The same software also permits to instruct the use of a number of threads that matches
optimally the amount of parallelism available on a multi-core target machine.

\section{Conclusions}\label{s:conc}
We presented
the properties of a family of algorithms 
which retain their structural characteristics though vary their operation depending on a
combinatorial parameter. We showed how such
``phenotypically plastic'' system~\cite{Jen04:Jen,DeF11a} may be used
to meet various changing requirements by tuning dynamically the
amount of algorithmic parallelism manifested by the software.
This paves the way to enabling robust control
on the emergence of several properties and behaviors, e.g.
collision avoidance, deterministic upper bounds on energy consumption,
and optimal use of the available communication resources.
Finally we showed how several evolution engines may be adopted to
achieve autonomic context-aware selection of members
that best match the current contextual conditions.

\bibliographystyle{latex8}

\end{document}